\documentclass[aps,prb,twocolumn,showpacs,floatfix,superscriptaddress,notitlepage]{revtex4-1}
\usepackage{dcolumn}
\usepackage{bm}
\usepackage{soul}
\usepackage{amsmath,amssymb,graphicx}
\usepackage[colorlinks=true,citecolor=blue,urlcolor=blue,linkcolor=blue]{hyperref}
\usepackage[export]{adjustbox}

\begin{document}

\title{Photoemission signature of excitons}

\author{Avinash Rustagi}
\email{arustag@ncsu.edu}
\affiliation{Department of Physics, North Carolina State University, Raleigh, NC 27695}

\author{Alexander~F.~Kemper}
\email{akemper@ncsu.edu}
\affiliation{Department of Physics, North Carolina State University, Raleigh, NC 27695}

\date{\today}
\begin{abstract}
Excitons - the particle-hole bound states - composed of localized electron-hole states in semiconducting systems are crucial to explaining the optical spectrum. Spectroscopic measurements can contain signatures of these two particle bound states and can be particularly useful in determining the characteristics of these excitons. We formulate an expression for evaluating the angle-resolved photoemission spectrum arising from the ionization of excitons given their steady-state distribution in a semiconductor. We show that the spectrum contains information about the direct/indirect band gap nature of the semiconductor and is located below the conduction band minimum displaced by the binding energy. The dispersive features of the spectrum contains remnants of the valence band while additional interesting features arise from different exciton distributions. Our results indicate that for most exciton probability distributions, the energy integrated photoemission spectrum provides an estimate of the exciton Bohr radius.
\end{abstract}
\maketitle

\section{Introduction}
\label{Introduction}
Excitons (bound states of electron-hole pairs) dominate the optical spectrum of a semiconductor at sub-band gap energies. Typically, they are observed using optical techniques such as absorption spectroscopy. They have gained significant recent interest due to recent observations of novel phenomena like exciton condensation \cite{kogar1314}, exciton-polariton condensates \cite{byrnes2014exciton}, and their possible applications in future optoelectronic devices \cite{xiao2017excitons}.

Although the optical techniques may yield significant information regarding the excitonic properties, the momentum-averaged nature of the probe may leave some aspects unanswered.  A different technique, angle-resolved photoemission spectroscopy (ARPES), is typically employed to provide complementary information, which has been quite successful in imaging the occupied electronic bands of materials \cite{luSC_Arpes,chenTI_Arpes}, discovering novel phases of matter like topological insulators \cite{xia2009TI} and Weyl semimetals \cite{xu2015WSM}, and studying origin of charge density waves (CDW) in 1T-TiSe$_2$ \citep{Cercellier_TiSe2_PRL_EI,Rossnagel_PRB_JT,Monney_TiSe2_PRB_EI}.  However, since it measures electrons from the occupied spectral function, it has no access to excitons by itself.  This may be remedied by employing a non-equilibrium version, time-resolved ARPES (tr-ARPES) that is capable of measuring the unoccupied states, as was dramatically demonstrated in topological insulator \cite{SobotaPRL_TI}. Tr-ARPES studies have led to a substantial progress in understanding carrier dynamics \cite{MoS2_CarrDyn2015}, band structure control \cite{Ultrafast_BandControl}, enhancing and tuning competeing ordered phases \cite{sentef_SC,sentef_SCCDW}, observing transient CDW gap melting in TbTe$_3$ \cite{Schmitt_Science2008}, and optical control of spin and valley polarization of excited carriers in WSe$_2$ \citep{Bertoni_PRL2016}. 

While the tools of many body theory have successfully explained the features in ARPES due to single-particle states, the signatures of excitons in ARPES requires further study. Recent studies have considered the formation of excitons transiently by the action of a pump pulse and measured their subsequent dynamics in metals \cite{petek2014transient,silkin2015ultrafast} and semiconductors \cite{weinelt2004dynamics,buss2017ultrafast}, thereby raising the question: What is the contribution of excitons to ARPES? 

Typically, to calculate the contribution of excitons to the ARPES requires coupling of the Bethe-Salpeter equation to the single particle Green function through self energy \cite{steinhoff2017exciton,perfetto2016first}. However, a significant step can be taken towards such a calculation assuming a steady-state distribution of excitons in the system. Although this approach ignores the means of creating long-lived excitons and the subsequent exciton-exciton interaction, it nevertheless elucidates their contribution to the measured spectra.

In most conventional bulk semiconductors, the dielectric screening of Coulomb interaction leads to small exciton binding energies $\mathcal{O}$(10 meV) \citep{SiGe_ExBE}, making it difficult to resolve such signatures constrained by energy resolution. However, there are systems hosting tightly bound excitons with large binding energies \cite{Baldini_TiO2,Rubio_BNNT} where such signatures can be significant. One such class of materials is the transition metal dichalcogenides (TMDCs) which in bulk (indirect gap semiconductors) are stacked planar layers, held weakly by Van der Waal forces. Thus, weakened screening of interaction lines lead to somewhat large exciton binding energies $\mathcal{O}$(100 meV). In addition, the monolayer TMDCs (direct gap semiconductors) with even weaker screening and quantum confinement has an even larger exciton binding energy $\sim$ 500 meV. This makes TMDCs ideal candidates to observe exciton signatures in photoemission measurements. While excitons in indirect gap semiconductors cannot be resonantly photoexcited due to violation of conservation of momentum, they may be indirectly formed via intervalley scattering \cite{MoS2_IV2ppe}. 

In this paper, we evaluate such a contribution and find that the spectroscopic signature from excitons appears below the conduction band minima displaced by the exciton binding energy. The location of the spectral signature in the Brillouin zone captures the direct/indirect gap nature of the bands, and the width of the spectrum in momentum is controlled by the size of the exciton through the wavefunction and the distribution function of excitons. In Sec.~\ref{Theory} of this paper, we describe our formalism for calculating the photoemission spectrum from excitons in a semiconductor. Results are presented and discussed in Sec.~\ref{Results and Discussion} and the conclusions of this paper are summarized in Sec.~\ref{Conclusion}.

\section{Theory}
\label{Theory}
\subsection{Exciton Creation Operator}
\label{Exciton Creation Operator}
\begin{figure}[htbp]
\centering
\includegraphics[width=\linewidth]{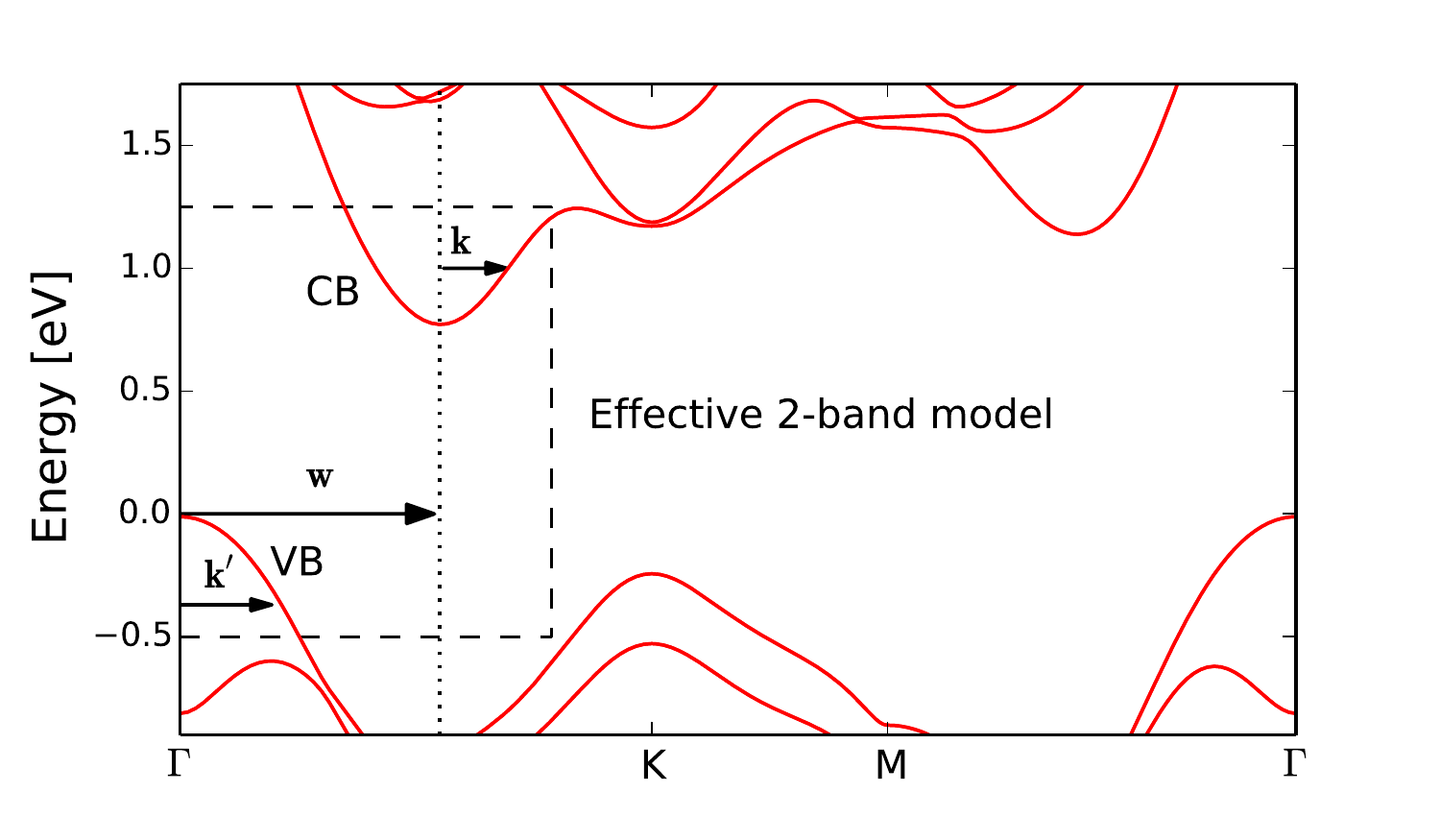}
\caption{\label{schematic} DFT electronic structure of the indirect-gap bulk semiconductor - MoSe$_2$, with the effective 2-band model highlighted (dashed box).}
\end{figure}

We consider a model two-band semiconductor within the effective mass approximation (see Fig.~\ref{schematic}) where the conduction band minima and the valence band maxima are separated by wavevector ${\bf w}$. To recover the case of a direct gap semiconductor, ${\bf w}$ may be set to zero. The Hamiltonian for such a model is 
%
%
\begin{equation}
\label{Hamiltonian}
\begin{split}
&H = H_0 + W \\
&H_0 = \sum_{\bm k'} \epsilon_{v,{\bm k'}}\, a^{\dagger}({\bm k'}) a({\bm k'}) +  \sum_{\bm k} \epsilon_{c,{\bm k}+{\bf w}} \, b^{\dagger}({\bm k}) b({\bm k})  \\
&W =\sum_{{\bm k},{\bm k'},{\bm q}\neq 0} V({\bm q}) b^{\dagger}({\bm k}+{\bm q}) a^{\dagger}({\bm k'}-{\bm q}) a({\bm k'}) b({\bm k})
\end{split}
\end{equation}
where $b^{\dagger}({\bm k})/b({\bm k})$ is the creation/annihilation operator for an electron in conduction band (CB) with momentum ${\bm k}+{\bf w}$ and $a^{\dagger}({\bm k'})/a({\bm k'})$ is the creation/annihilation operator for an electron in valence band (VB) with momentum ${\bm k'}$. The dispersion for the bands are (we use natural units and set $\hbar=1$) 
\begin{equation}
\epsilon_{v,{\bm k'}}= -\dfrac{\bm k'^2}{2 m_v} \qquad \epsilon_{c,{\bm k}}= E_g + \dfrac{({\bm k}-{\bf w})^2}{2 m_c}
\end{equation}
where $m_c/m_v$ is the effective mass for the CB/VB and energy gap $E_g$. Following the seminal work on excitons by Elliott \cite{Elliott1957}, the exciton creation operator with center of mass (COM) momentum ${\bm Q}$ and state $\lambda$ can be written as a superposition of the electron-hole pair operators
\begin{equation}
\label{ExcitonCreation}
A_{\lambda}^\dagger ({\bm Q},{\bf w}) = \sum_{\bm p} \phi^{{\bf w}}_{\lambda}({\bm p}) b^{\dagger}({\bm p}+\alpha {\bm Q}) a({\bm p}-\gamma {\bm Q})
\end{equation}
where $\alpha=m_c/M$, $\gamma=m_v/M$, $M=m_c+m_v$, and $ \phi^{{\bf w}}_{\lambda}({\bm p})$ is the envelope wavefunction which satisfies the eigenvalue equation 
\begin{equation}
\begin{split}
\left[ \dfrac{p^2}{2 \mu} -E_{\lambda}\right] \phi^{{\bf w}}_{\lambda}({\bm p}) &= \sum_{{\bm p'}} V({\bm p}-{\bm p'}) \phi^{{\bf w}}_{\lambda}({\bm p'})
\end{split}
\end{equation}
with $\mu^{-1}=m_c^{-1}+m_v^{-1}$ being the reduced mass of the electron-hole pair and $E_{\lambda}$ being the eigenvalue \cite{clerc2006photoemission}. The equation of motion for the exciton creation operator is given by 
\begin{widetext}
\begin{equation}
\begin{split}
i \dfrac{\partial}{\partial t} A^\dagger ({\bm Q},{\bf w}) \vert G \rangle &= \sum_{\bm p} \phi^{{\bf w}}_{\lambda}({\bm p}) \left[ i \dfrac{\partial}{\partial t} b^{\dagger}({\bm p}+\alpha {\bm Q}) a({\bm p}-\gamma {\bm Q}) + b^{\dagger}({\bm p}+\alpha {\bm Q}) i \dfrac{\partial}{\partial t} a({\bm p}-\gamma {\bm Q}) \right] \vert G \rangle \\
&= \sum_{\bm p} \left[ \epsilon_{v,{\bm p}-\gamma {\bm Q}} - \epsilon_{c,{\bm p}+\alpha {\bm Q} +{\bf w}}\right] \phi^{{\bf w}}_{\lambda}({\bm p}) b^{\dagger}({\bm p}+\alpha {\bm Q}) a({\bm p}-\gamma {\bm Q}) \vert G \rangle\\
& \quad + \sum_{{\bm k'},{\bm p},{\bm q}\neq 0} V({\bm q}) \phi^{{\bf w}}_{\lambda}({\bm p})  b^{\dagger}({\bm p}+\alpha {\bm Q}+{\bm q}) a^\dagger({\bm k'}-{\bm q}) a({\bm p}-\gamma {\bm Q}) a({\bm k'}) \vert G \rangle\\  
& \quad + \sum_{{\bm k},{\bm p},{\bm q}\neq 0} V({\bm q})  \phi^{{\bf w}}_{\lambda}({\bm p}) b^{\dagger}({\bm p}+\alpha {\bm Q}) a({\bm p}-\gamma {\bm Q}+{\bm q})  b^{\dagger}({\bm k}+{\bm q}) b({\bm k}) \vert G \rangle
\end{split}
\end{equation}
\end{widetext}
Considering the ground state $\vert G \rangle $ to be completely filled VB and empty CB implies
\begin{equation}
b({\bm k}) \vert G \rangle =0  \qquad  a^\dagger ({\bm k}) \vert G \rangle =0 
\end{equation}
and accounting for the absence of $q=0$ term in the sum, we can rearrange the Coulomb interaction terms using the Fermionic anti-commutation rules to get
\begin{equation}
\begin{split}
i \dfrac{\partial}{\partial t} A^\dagger ({\bm Q},{\bf w})&= \left[ -E_g - E_{\lambda} - \dfrac{{\bm Q}^2}{2M}\right] A^\dagger ({\bm Q},{\bf w}) 
\end{split}
\end{equation}
The eigenvalue $E_\lambda =-E^{B}_{\lambda}< 0$ where $E^{B}_{\lambda}$ is the exciton binding energy. Therefore an exciton with COM momentum ${\bm Q}$ in state $\lambda$ has energy 
\begin{equation}
E_{\lambda,{\bm Q}} = E_g - E^{B}_\lambda + \dfrac{{\bm Q}^2}{2 M}
\end{equation}
\subsection{ARPES}
\label{ARPES}
To evaluate the ARPES expression, we follow the semi-perturbative theoretical description developed by Freericks et. al. to consider the quasi-equilibrium situation where excitons exist in the system \cite{freericks_ARPES}. The system is described in the distant past ($t\rightarrow -\infty$) by the Hamiltonian $H$ and the many-body eigenstates characterized by $E_{n}$ and $\vert \Psi_{n}\rangle$: 
\begin{equation}
H \vert \Psi_{n}\rangle = E_{n} \vert \Psi_{n}\rangle. 
\end{equation}
The action of the pump modifies the Hamiltonian to $H_{pump}(t)$ and the eigenstates at $t_0$ is given by 
\begin{equation}
\vert \Psi_{n}^{I} (t_0)\rangle = U(t_0,-\infty) \vert \Psi_{n}\rangle
\end{equation}
where the time-evolution operator
\begin{equation}
\begin{split}
U(t,t_0) &= \hat{T}_t  \exp \left( -i \int_{t_0}^{t} dt_1 H_{pump}(t_1) \right)
\end{split}
\end{equation}
With the addition of the probe, the Hamiltonian is $H_{pump}(t) + H_{probe}(t)$ and the eigenstates at $t$ is given by 
\begin{equation}
\vert \Psi_{n}^{F} (t)\rangle = \bar{U}(t,t_0) \vert \Psi_{n}^{I}(t_0)\rangle
\end{equation}
where the time-evolution operator $\bar{U}(t,t_0)$ is 
\begin{equation}
\bar{U}(t,t_0) = \hat{T}_t \exp \left( -i \int_{t_0}^{t} dt_1\left[ H_{pump}(t_1)+H_{probe}(t_1)\right] \right)
\end{equation}
In the limit of weak probe, $H_{probe}$ can be treated perturbatively and the time-evolution operator can be linearized
\begin{equation}
\label{linearized_U}
\bar{U}(t,t_0) \simeq U(t,t_0) -i \int_{t_0}^{t} dt_1 U(t,t_1) H_{probe}(t_1) U(t_1,t_0)
\end{equation}
The probability to find a photoelectron with momentum $k$ in an interval $dk$ in solid angle $d\Omega_{k}$ is
\begin{equation}
\label{expr_P}
\begin{split}
& I(t)=\lim_{t\rightarrow \infty}\dfrac{k^2 dk d\Omega_{k}}{(2\pi)^3} P(t); P(t)=\sum_{n,m} \rho_n \vert \langle \Psi_{m}^{1<}; {\bm k} \vert \Psi_{n}^{F}(t)\rangle\vert^2 \\
\end{split}
\end{equation}
$\rho_n$ is the density matrix corresponding to occupation of the state $n$. The final state after the action of the probe Hamiltonian denoted by $\vert \Psi_{m}^{1<}; {\bm k} \rangle \equiv \vert \Psi_{m}^{1<} \rangle \otimes \vert {\bm k} \rangle$ has a state with one lesser electron and a free photoemitted electron. The photoemission probability expression in Eq.~\ref{expr_P} implies that the initial state for the photoemission can be any from the ensemble of initial states $\vert \Psi_{n}\rangle$ with probability $\rho_n$ that is time-evolved by the evolution operators whereas $\vert \Psi_{m}^{1<} \rangle$ is the state the system is left in after photoemission of an electron i.e. state with one lesser electron. 

The probe Hamiltonian term describing photoemission from an exciton state due to a probe pulse with energy $\omega_0$ and temporal profile $s(t)$ is 
\begin{equation}
H_{probe} (t) = s(t) e^{-i\omega_0 t} M^{f c}_{{\bm k},{\bm k}^\prime} f^{\dagger}({\bm k}) b({\bm k}^{\prime}-{\bf w})
\end{equation}
which annihilates a CB electron with momentum ${\bm k}$ and creates a free electron $f^\dagger$ with matrix element $M^{f c}_{{\bm k},{\bm k}^\prime}$. Note that $b({\bm k}^\prime-{\bf w})$ annihilates an electron from the CB with momentum ${\bm k'}$. The process of photoemission conserves the parallel component of momenta but not the perpendicular component. Thus the probe creating a free electron with momentum ${\bm k}$ and annihilating a CB electron with momentum ${\bm k}^\prime$ should conserve the parallel component of momentum. This implies that ${\bm k}=\{k_{||},k_z \}$ and ${\bm k}^\prime=\{k_{||},k_z^\prime \} $. However, this still leaves an uncertainty with respect to the z-component of momentum. The z-component can empirically be chosen by varying the photon energy, and this is often employed to measure the c-axis dispersion. However, there is always an unknown offset since the z-component is not conserved. For this reason, we set $k_z^\prime =0$ but we note that for a fixed probe photon energy, we might have used any finite value for $k_z^\prime$ given the unknown offset. Using the linearized time-evolution operator $\bar{U}(t,t_0)$ (Eq.~\ref{linearized_U}), the required matrix element is therefore evaluated to be
\begin{widetext}
\begin{equation}
\begin{split}
\langle \Psi_{m}^{1<}; {\bm k} \vert \Psi_{n}^{F}(t)\rangle &\simeq  -i \int_{t_0}^{t} dt_1 \langle \Psi_{m}^{1<}; {\bm k} \vert  U(t,t_1) H_{probe}(t_1) U(t_1,t_0) \vert \Psi_{n}^{I}(t_0)\rangle  \\
&\simeq  -i \int_{t_0}^{t} dt_1 s(t_1) e^{-i\omega_0 t_1} e^{-i\omega_e (t-t_1)} M^{f c}_{{\bm k},{\bm k}^\prime}\langle \Psi_{m}^{1<} \vert  U(t,t_1) b({\bm k}^\prime -{\bf w}) U(t_1,t_0)  \vert \Psi_{n}^{I}(t_0)\rangle \\
\end{split}
\end{equation}
\end{widetext}
which therefore implies
\begin{widetext}
\begin{equation}
\begin{split}
\sum_m \vert \langle \Psi_{m}^{1<}; {\bm k} \vert \Psi_{n}^{F}(t)\rangle\vert^2 &\simeq  \int_{t_0}^{t} dt_1  \int_{t_0}^{t} dt_2 s(t_1) s(t_2) e^{i(\omega_0 -\omega_e-W) (t_2- t_1)}  \vert M^{fc}_{{\bm k},{\bm k}^\prime} \vert^2 \\
& \times   \langle \Psi_{n}^{I}(t_0) \vert  U(t_0,t_2) b^\dagger({\bm k}^\prime -{\bf w}) U(t_2,t) \underbrace{\sum_m \vert \Psi_{m}^{1<} \rangle \langle \Psi_{m}^{1<}}_{=1} \vert  U(t,t_1) b({\bm k}^\prime -{\bf w}) U(t_1,t_0)  \vert \Psi_{n}^{I}(t_0)\rangle \\
\end{split}
\end{equation}
\end{widetext}
where $\vert \Psi_m^{1<} \rangle$ (eigenstates of the system with one lesser electron i.e. the hole state) forms a complete basis set, $\omega_e$ is the kinetic energy of photoemitted electron and $W$ is the work function of the material. 
%
%
\begin{figure}[h!]
\centering
\includegraphics[width=\linewidth, frame]{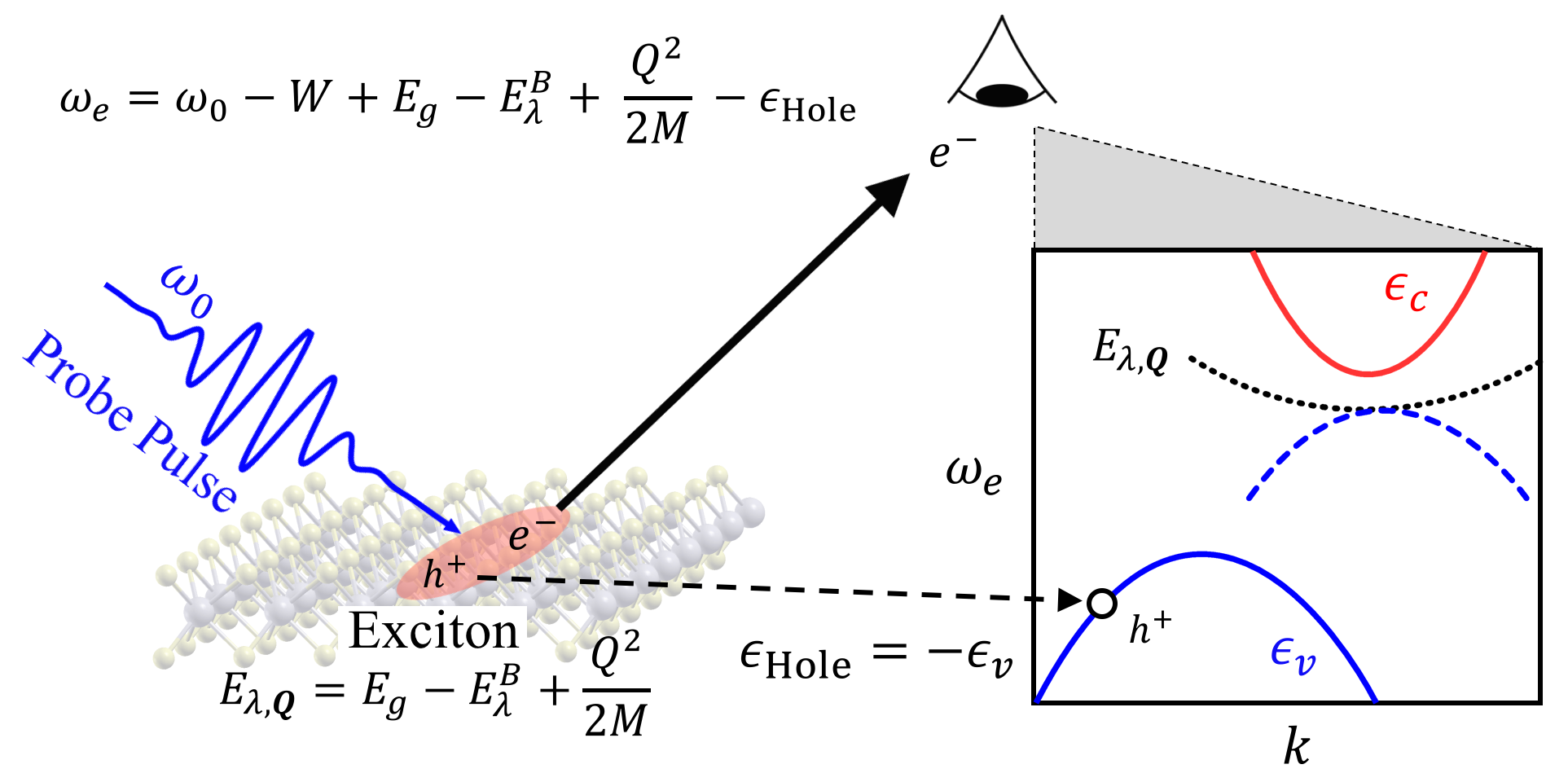}
\caption{\label{cartoon} A schematic showing the photoemission process and the corresponding energy conservation rule.}
\end{figure}
%
%

The exciton excitations are eigenstates of the interacting electron-hole Hamiltonian in Eq.~\ref{Hamiltonian} with eigenstates characterized by the index $n=\{\lambda,{\bm Q}\}$. To evaluate the spectral signatures of excitons in ARPES, we consider the regime in which the pump has created excitons which now are in a steady-state. We note that in assuming the existence of excitons in the system, our approach does not capture the information on the early time coherent dynamics of excited particles by the pump. In our approach, where the pump is over and has created excitons in a steady-state serves as a starting point such that $\vert \Psi_{n}^{I}(t_0)\rangle \equiv \vert \Psi_{\lambda,{\bm Q}}\rangle$ and $H_{pump}(t)=H$, 
\begin{equation}
\begin{split}
H &\vert \Psi_{\lambda,{\bm Q}}\rangle = \left(\Omega_0+E_{\lambda,{\bm Q}}\right) \vert \Psi_{\lambda,{\bm Q}}\rangle \\
\Rightarrow U(t,t')  &\vert \Psi_{\lambda,{\bm Q}}\rangle = e^{-i \left(\Omega_0+E_{\lambda,{\bm Q}}\right)(t-t')} \vert \Psi_{\lambda,{\bm Q}}\rangle
\end{split}
\end{equation}
where $\Omega_0=\sum_{\bm k} \epsilon_{v,{\bm k}}$ is the ground state energy of the system (filled VB and empty CB). We can thus evaluate the probability in Eq.~\ref{expr_P}, 
\begin{equation}
\begin{split}
P(t) &\simeq \int_{t_0}^{t} dt_1 \int_{t_0}^{t} dt_2 s(t_1) s(t_2) e^{i(\omega_0 -\omega_e-W) (t_2- t_1)} \\
&\times \vert M^{fc}_{{\bm k},{\bm k}^\prime} \vert^2 \sum_{\lambda, {\bm Q}} \rho_{\lambda,{\bm Q}} e^{i \left(\Omega_0+E_{\lambda,{\bm Q}}\right)(t_2-t_1)}\langle \Psi_{\lambda,{\bm Q}}\vert \\
&\times   b^\dagger ({\bm k}^\prime-{\bf w}) U(t_2,t_1) b({\bm k}^\prime-{\bf w})  \vert \Psi_{\lambda,{\bm Q}}\rangle
\end{split}
\end{equation}
Here we have used the \textit{sudden} approximation (valid only when photoelectron energies are high) where the removal of photoelectron is instantantaneous \cite{Damascelli_RMP2003}. We can expand the exciton state using the exciton creation operator acting on the ground state: $\vert \Psi_{\lambda,{\bm Q}}\rangle =A_\lambda^\dagger ({\bm Q},{\bf w})\vert G\rangle$ 
\begin{equation}
\begin{split}
&U(t_2,t_1) b({\bm k}^\prime-{\bf w})  \vert \Psi_{\lambda,{\bm Q}}\rangle = e^{-i\left(\Omega_0-\epsilon_{v,{\bm k}^\prime-{\bm Q}-{\bf w}}\right)(t_2-t_1)} \\
& \qquad \times \phi^{{\bf w}}_{\lambda}({\bm k}^\prime-\alpha {\bm Q} - {\bf w}) a({\bm k}^\prime- {\bm Q}-{\bf w}) \vert G\rangle
\end{split}
\end{equation}
%
%
\begin{figure}[h!]
\includegraphics[width=\linewidth]{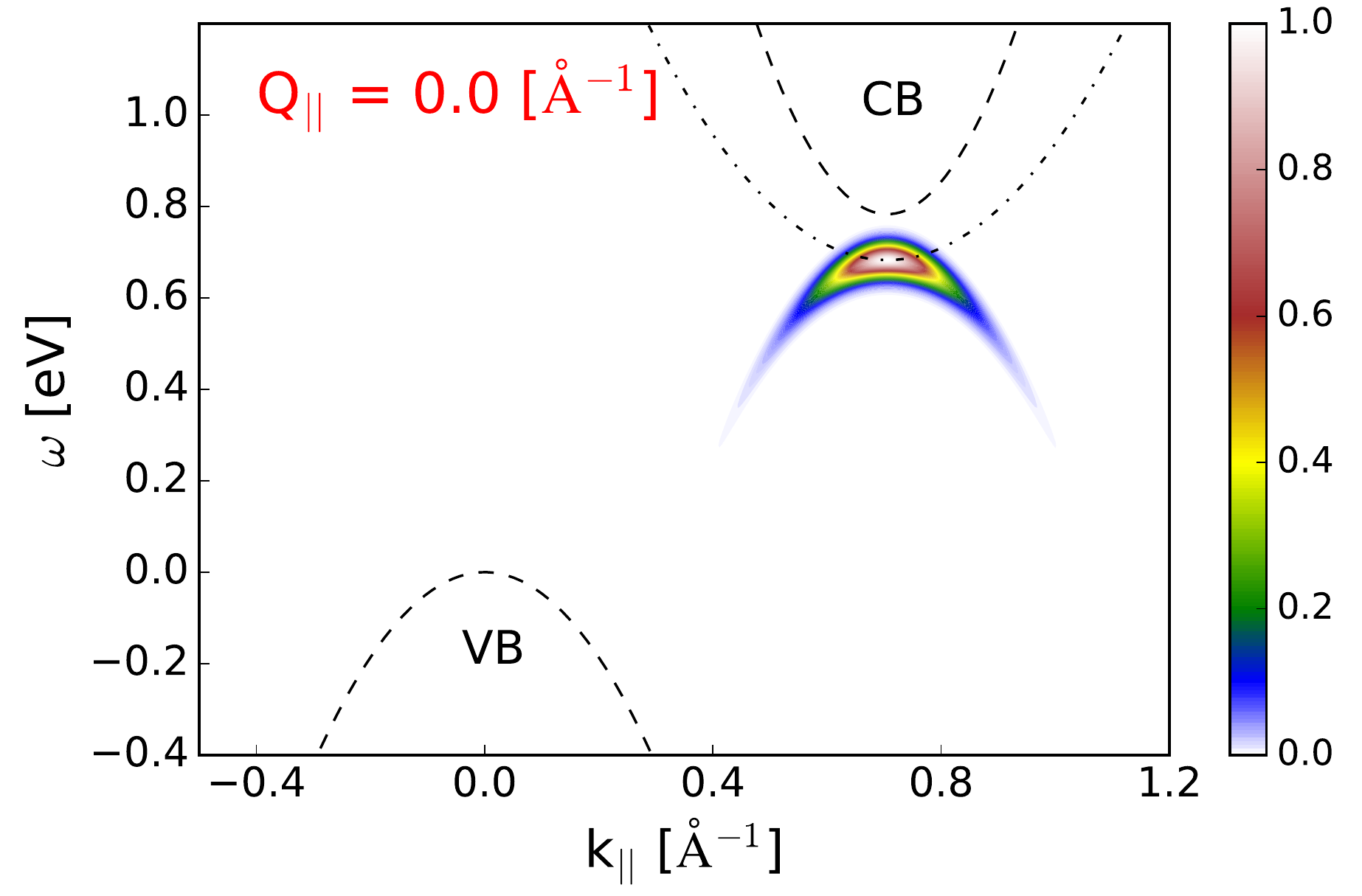}
\caption{\label{ZeroT_ZeroQ} ARPES spectra for ${\bm Q}=0$ exciton. The VB dispersion is mapped by the photoemitted electron $\{\omega_e,k_{||}\}$.}
\end{figure}
%
%
%
%
\begin{figure*}[htbp]
\includegraphics[width=\linewidth]{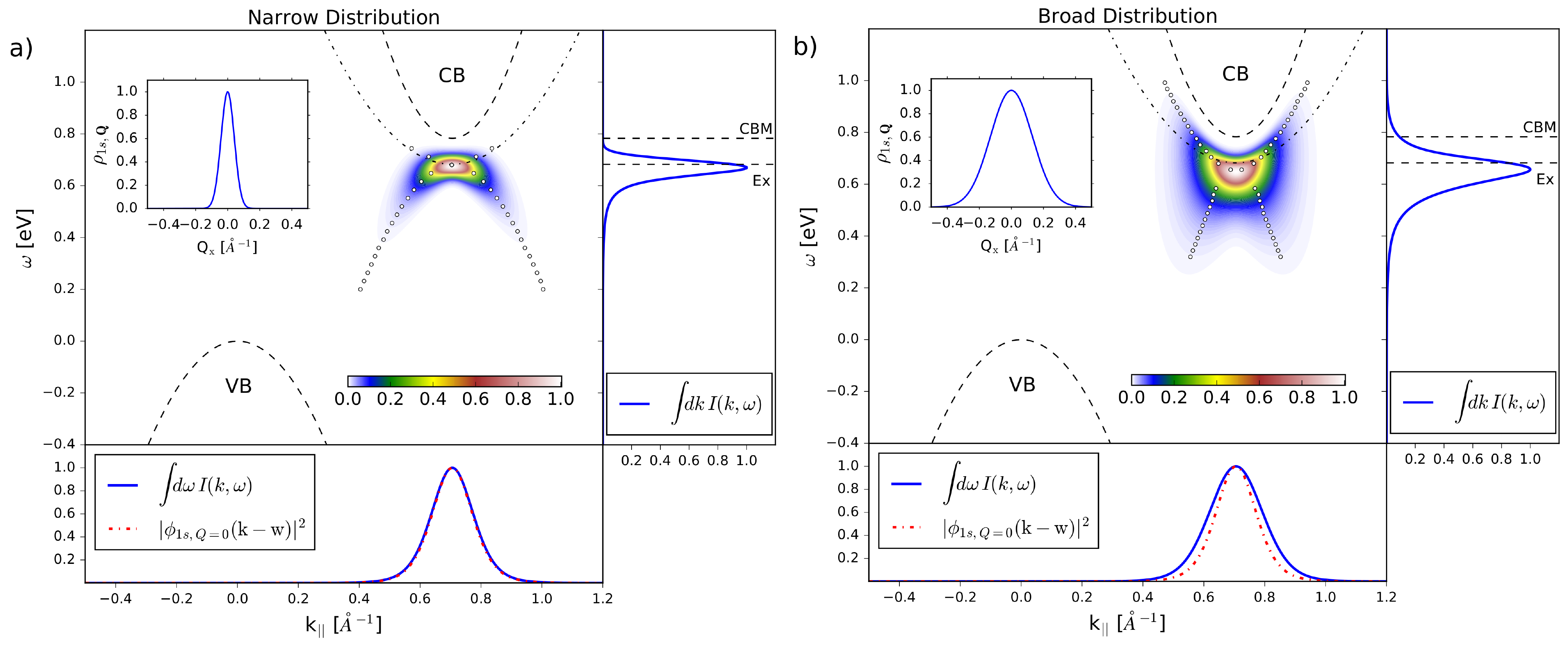}
\caption{\label{Arpes_Spectra} ARPES spectra for a) narrow ($\beta$= 100 eV$^{-1}$), and b) broad exciton distribution ($\beta$= 10 eV$^{-1}$) (shown in inset) in a model two-band indirect gap semiconductor with 1s-exciton state. Dispersive features are highlighted by empty circles. The energy integrated spectra compared to exciton wavefunction squared is below the panels and the momentum integrated spectra is to the right. }
\end{figure*}
%
%
%
%
\begin{figure}[htbp]
\centering
\includegraphics[width=\linewidth]{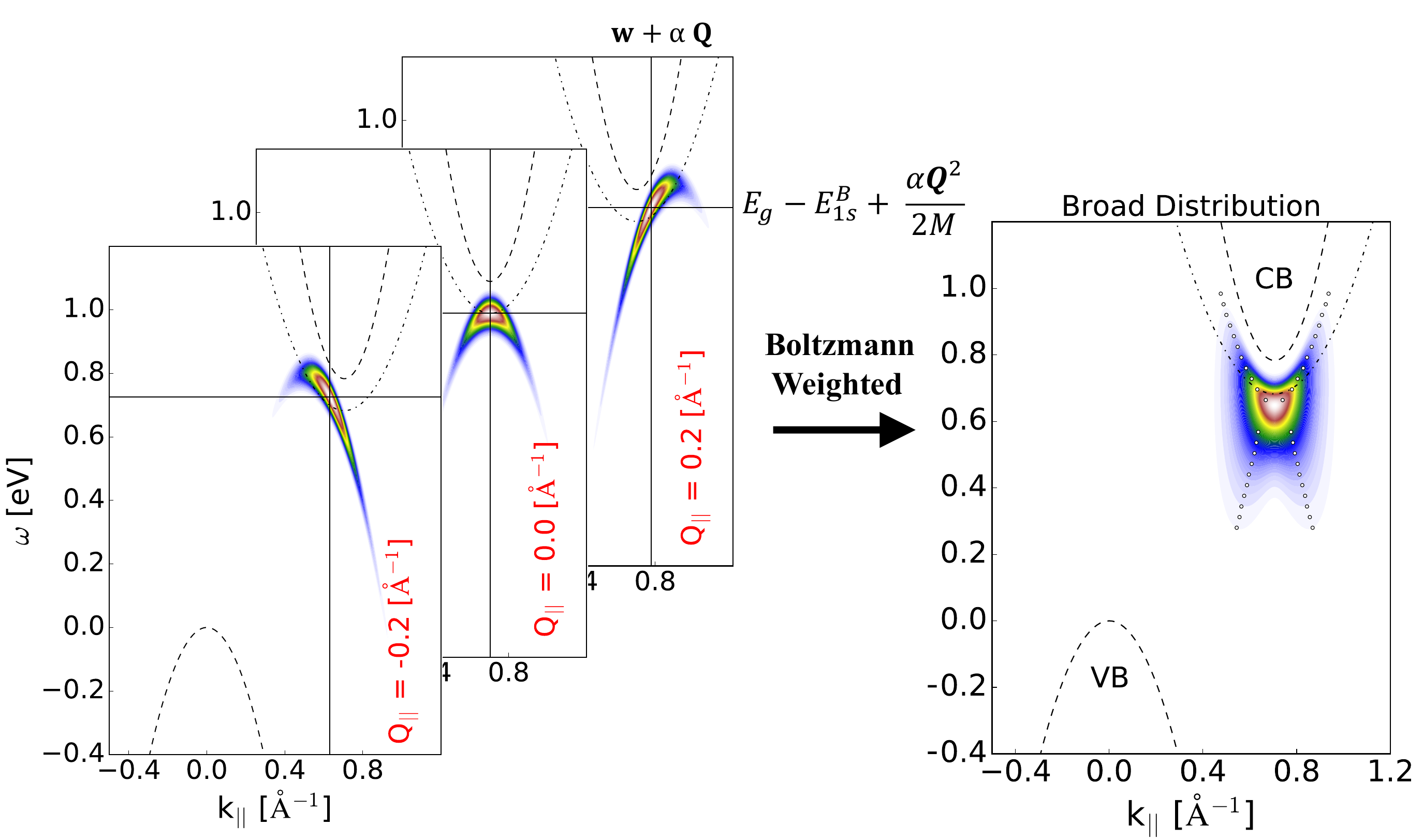}
\caption{\label{boltzmann} The contributions from different ${\bm Q}$-excitons are Boltzmann weighted to get the complete spectra. }
\end{figure}
%
%
The state $ a({\bm k}^\prime- {\bm Q}-{\bf w}) \vert G\rangle $ has one lesser electron in the VB with energy $\Omega_0-\epsilon_{v,{\bm k}^\prime-{\bm Q}-{\bf w}}$. Assuming Gaussian probe pulse envelope for $s(t)$ centered about delay time $t_d$ with temporal width $\sigma$, setting $t_0 \rightarrow -\infty$ (given that the probe delay time $t_d$ is at later times when the pump is over and the created excitons are in a steady-state) and taking the long time limit $t \rightarrow \infty$, 
\begin{equation}
\begin{split}
P(t) &\simeq \int_{-\infty}^{\infty} dt_1 \int_{-\infty}^{\infty} dt_2 e^{-\dfrac{(t_1-t_d)^2}{2 \sigma^2}} e^{-\dfrac{(t_2-t_d)^2}{2 \sigma^2}}  \\
&\times  e^{i(\omega_0 -\omega_e-W) (t_2- t_1)} \vert M^{fc}_{{\bm k},{\bm k}^\prime} \vert^2 \sum_{\lambda, {\bm Q}} \rho_{\lambda,{\bm Q}} e^{i E_{\lambda,{\bm Q}}(t_2-t_1)}\\
&\times  e^{i \epsilon_{v,{\bm k}^\prime-{\bf w}-{\bm Q}}(t_2-t_1)}  \vert \phi^{{\bf w}}_{\lambda}({\bm k}^\prime-{\bf w}-\alpha {\bm Q})\vert^2 .
\end{split}
\end{equation}
Wigner transforming the time variables to the average and relative time $[t_1,t_2]\rightarrow \left[ t_a=(t_1+t_2)/2; t_r=t_1-t_2\right]$ and integrating, 
\begin{equation}
\label{finalexpr_P}
\begin{split}
P(t_d) &\simeq 2\pi \sigma^2 \vert M^{fc}_{{\bm k},{\bm k}^\prime} \vert^2  \sum_{\lambda, {\bm Q}} \rho_{\lambda,{\bm Q}} \vert \phi^{{\bf w}}_{\lambda}({\bm k}^\prime-{\bf w}-\alpha {\bm Q})\vert^2 \\
&\times  \exp \left(-\sigma^2 [-\omega+E_{\lambda,{\bm Q}}+\epsilon_{v,{\bm k}^\prime-{\bf w}-{\bm Q}}]^2\right) 
\end{split}
\end{equation}
where $\omega=\omega_e+W-\omega_0$ is the difference in energy of the material before and after photoemission. Eq.~\ref{finalexpr_P} depends on the exciton wavefunction, exciton distribution, and energy conservation. Integrating the ARPES intensity over momentum provides information of the location of the signal in energy while integrating the intensity over energy displays the spread of signal in momentum. The energy integrated photoemission intensity is 
\begin{equation}
\begin{split}
\int d\omega P(t_d) &\simeq  2\pi^{3/2} \sigma \vert M^{fc}_{{\bm k},{\bm k}^\prime} \vert^2 \sum_{ \lambda, {\bm Q}} \rho_{\lambda,{\bm Q}} \vert \phi^{{\bf w}}_{\lambda}({\bm k}^\prime-{\bf w}-\alpha {\bm Q})\vert^2   
\end{split}
\end{equation}
%
%
\begin{figure*}[htbp]
\includegraphics[width=\linewidth]{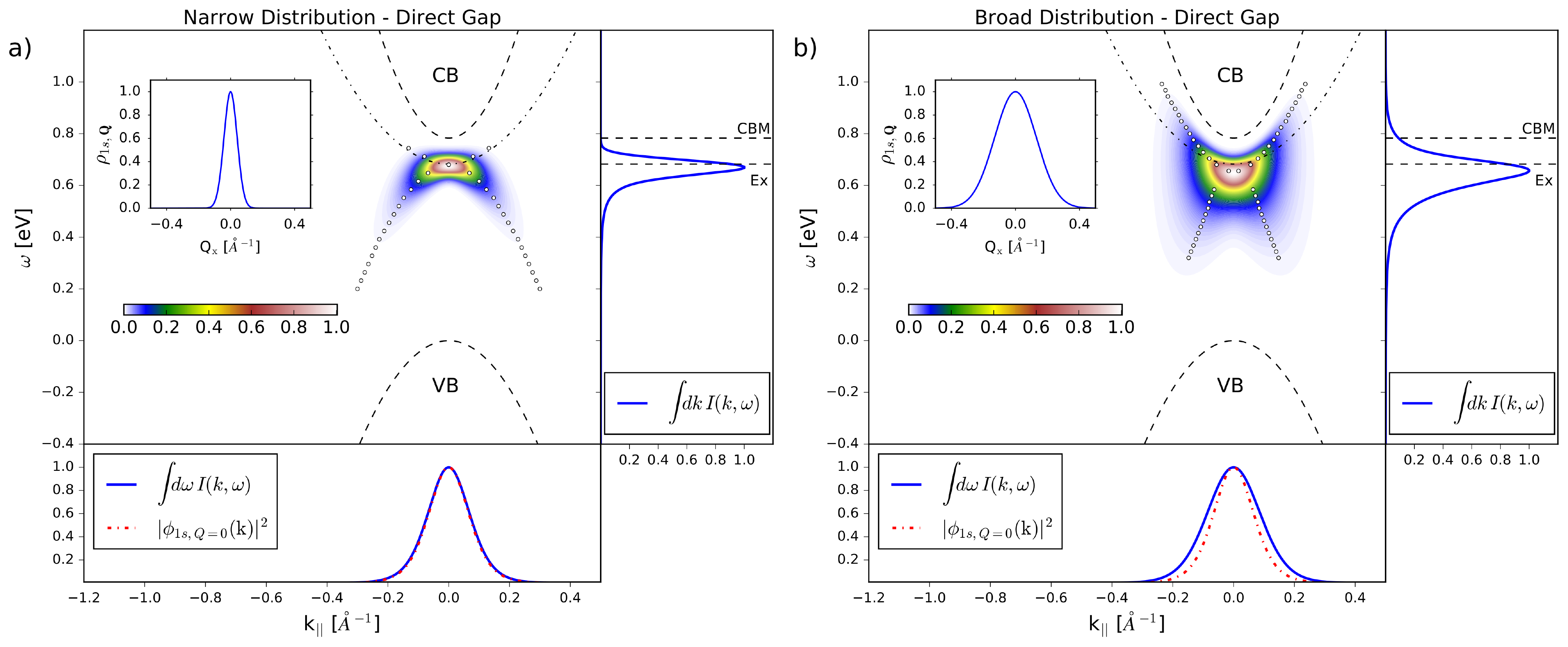}
\caption{\label{Arpes_Spectra_Direct}  ARPES spectra for a) narrow ($\beta$= 100 eV$^{-1}$), and b) broad exciton distribution ($\beta$= 10 eV$^{-1}$) (shown in inset) in a model two-band direct gap semiconductor with 1s-exciton state. Dispersive features are highlighted by empty circles. The energy integrated spectra compared to exciton wavefunction squared is below the panels and the momentum integrated spectra is to the right.}
\end{figure*}
%
%
\section{Results and Discussion}
\label{Results and Discussion}
Our expression for photoemission $P(t_d)$ is based on the quasi-equilibrium steady-state situation where excitons exist described by the Boltzmann weight with an effective exciton temperature describing the distribution of excitons with finite COM momentum, $\rho_{\lambda,{\bm Q}} \propto \exp\left[-\beta (E_{\lambda,{\bm Q}}-E_G+E^{B}_{\lambda})\right]$. Larger the exciton temperature (smaller $\beta$), higher COM momentum states are occupied i.e. broad distribution and vice-versa. We emphasize that the exciton temperature is not a measure of temperature in the conventional sense but is a measure of the distribution of excitons.

We apply our theory to the model case of hydrogenic 1s-excitons ($\lambda$=1s) in a two-band indirect gap semiconductor within the effective mass approximation with probe temporal width $\sigma=$ 20 fs. We assume the matrix element $M^{fc}_{{\bm k},{\bm k}^\prime}$ to be momentum independent. We use the parameters for bulk MoSe$_2$: $m_c=0.4794$, $m_v=0.8184$ in terms of the free electron mass, energy gap $E_g=0.7830$ eV, effective dielectric constant $\epsilon=6.4$ chosen such that the 1s-exciton binding energy matches the experimental exciton Rydberg $E^{B}_{\lambda=1s}=R_\mathrm{ex}=13.6056 \mu/\epsilon^2$ eV = 0.1 eV \cite{ANEDDA_MoSe2Exciton}. The corresponding `1s' wavefunction is $\phi^{{\bf w}}_{1s}({\bm p}) \propto 1/[1+p^2 a_\mathrm{ex}^2/4]^2$ ($a_\mathrm{ex}=0.529 \epsilon/\mu \, \mathrm{ \AA}$: exciton Bohr radius). We can also treat the case of a direct gap semiconductor by setting ${\bf w}=0$.

The photoemission process can be depicted by a diagram shown in Fig.~\ref{cartoon} which highlights the signal from an exciton in state $\lambda$ with COM momentum ${\bm Q}$. The initial energy (=$\omega_0+E_{\lambda,{\bm Q}}$) includes the probe photon and the exciton. The final energy (=$\omega_e+W+\epsilon_{\text{Hole}}$) stems from the ionization of exciton creating a photoemitted electron with kinetic energy $\omega_e$, which escapes the material work function $W$ leaving a hole with energy $\epsilon_{\text{Hole}}=-\epsilon_{v}$. The photoemitted electron $\{\omega_e,k\}$ maps the signal from an exciton characterized by $(\lambda,{\bm Q})$ to the VB dispersion about an energy which maximizes the product $\vert \phi^{{\bf w}}_{\lambda}({\bm k}^\prime-{\bf w}-\alpha {\bm Q})\vert^2 \exp \left(-\sigma^2 [-\omega+E_{\lambda,{\bm Q}}+\epsilon_{v,{\bm k}^\prime-{\bf w}-{\bm Q}}]^2\right)$. The exciton being a linear superposition of e-h pairs (see Eq.~\ref{ExcitonCreation}) allows for the photoemitted electron to have a range of momenta which manifests as a VB dispersion over a momentum range set by the exciton wavefunction in ARPES. This is most easily seen in Fig.~\ref{ZeroT_ZeroQ} where the spectra for ${\bm Q}=0$ 1s-exciton is evaluated which corresponds to $\beta \rightarrow \infty$. 

To display the effects of the presence of different COM momenta excitons, the spectra is evaluated for two exciton distributions: narrow ($\beta=100$ eV$^{-1}$ - Fig.~\ref{Arpes_Spectra}a)  and broad ($\beta=10$ eV$^{-1}$ - Fig.~\ref{Arpes_Spectra}b ). The inset in each of the ARPES false-color plot shows the exciton distribution. The conduction and valence bands are shown (dashed lines) while the exciton dispersion, which in principle lives in the two-particle excitation space is also shown (dash-dotted line) to act as a guiding tool. A significant feature of the spectra is that the exciton contribution lies below the conduction band minima (CBM) separated by the exciton binding energy as seen in the momentum integrated intensity plots to the right. The energy integrated ARPES spectra are plotted below each ARPES shows that for narrow exciton distribution, ${\bm Q}=0$ excitons are predominantly present and thus the energy integrated spectra overlaps with the exciton wavefunction $\vert \phi_{1s}({\bm k}^\prime-{\bf w})\vert^2$. Even in the case for broad distribution,  the energy integrated spectra closely follows the exciton wavefunction squared. Thus the exciton ARPES can be a valuable tool to estimate the exciton binding energy/Bohr radius (momentum/energy integrated spectra). This is consistent with an earlier study by Ohnishi et. al. treating a single $\Gamma$-point and saddle point exciton in GaAs\cite{Nasu_2017Exciton}. However, we have shown that such an inference holds even for the case of a broad exciton distribution. 

It is particularly interesting to look at the dispersive features of the photoemission spectrum. The empty circles in the ARPES spectra (Fig.~\ref{Arpes_Spectra}a and \ref{Arpes_Spectra}b) denote the dispersive features of the spectra evaluated using the minimum gradient method \citep{mingradient2017}. For the narrow exciton distribution which predominately has ${\bm Q}=0$ excitons, we see the VB dispersion located at energy $E_{1s,{\bm Q}=0}$ over a momentum range set by the exciton wavefunction. In contrast, the broad distribution case is rich showing a trace similar to the CB ($\omega > E_g-E^B_{1s}$) and also some remnants of VB-like dispersion ($\omega < E_g-E^B_{1s}$). Fig.~\ref{boltzmann} explains the origin of such features by evaluating the contribution of different COM momentum ${\bm Q}$ excitons. For each of the ${\bm Q}$ excitons, a contribution of the VB dispersion comes about the local extrema in momentum and energy. The exciton wavefunction has a maxima at the wavevector ${\bm k}^\prime={\bf w}+\alpha {\bm Q}$. Thus the energy about which the VB dispersion comes about is $E_g-E^{B}_{1s}+\alpha {\bm Q}^2/2M$. The mass ratio $\alpha=m_c/M$ makes the location of the ARPES contribution from ${\bm Q}$-exciton at an energy different from the exciton dispersion energy. The narrow exciton distribution predominantly has ${\bm Q}=0$ excitons and thus has a copy of the VB dispersion at energy $\omega=E_{1s,{\bm Q}=0}$. However, for the broad exciton distribution, the Boltzmann weighted contributions of VB dispersion from each of the ${\bm Q}$-excitons shows dispersion at $E_g-E^{B}_{1s}+\alpha {\bm Q}^2/2M$ and also remnants for the energies below the exciton dispersion minima. The Boltzmann weighted spectra over the exciton distribution is responsible for the lowering of the ARPES intensity below the exciton dispersion minima where the amount of lowering is set by the distribution width. The same formalism can be applied to the case of direct gap semiconductor (${\bf w}=0$) having a narrow and broad distribution of excitons seen in Fig.~\ref{Arpes_Spectra_Direct}. The conclusions drawn from the case of direct gap semiconductor are the same as indirect gap semiconductor.

\section{Conclusion}
\label{Conclusion}
In this paper, we have formulated an expression for evaluating the photoemission intensity from steady-state excitons which applies to indirect/direct gap semiconductors. The theory is applied to 1s-excitons in indirect gap bulk MoSe$_2$ within the effective mass approximation. The signal captures the indirect gap nature of the semiconductor seen in the location of the exciton contribution in the Brillouin zone and lies below the conduction band minima displaced by the exciton binding energy seen clearly in momentum integrated spectra. The energy integrated ARPES spectra can provide a close estimate of the exciton size irrespective of the exciton steady-state distribution. In addition, interesting dispersive features in the photoemission spectrum arises from the different exciton distributions. \\

\section{Acknowledgments}
\label{Acknowledgments}
The authors acknowledge the discussions and insights from Robert Kaindl, Martin Wolf, Ralph Ernstorfer and Laurenz Rettig.

\bibliography{ref}

\end{document}